\documentclass{PoS}
\usepackage{amssymb}
\usepackage{bm}
\usepackage{latexsym}

\newcommand{\possy}[2]{\href{http://pos.sissa.it/cgi-bin/reader/contribution.cgi?id=#1}{\tt #2}}

\newcommand{\Fone}{\ensuremath{\mathcal{F}(1)}}
\newcommand{\Vcb}{\ensuremath{V_{\mathrm{cb}}}}

\newcommand{\case}[2]{\ensuremath{\textstyle\frac{#1}{#2}}}
\newcommand{\cSW}{\ensuremath{c_{\mathrm{SW}}}}
\newcommand{\kch}{\ensuremath{\kappa_{\mathrm{c}}}}
\newcommand{\kbot}{\ensuremath{\kappa_{\mathrm{b}}}}

\title{$\bm{B\to D^{*}l\nu}$ at zero recoil: an update}

\ShortTitle{$B \to D^*l\nu$}

\author{
Jon~A.~Bailey,$^a$
A.~Bazavov,$^b$ 
C.~Bernard,$^c$ 
C.M.~Bouchard,$^{a,d}$ 
C.~DeTar,$^e$ 
A.X.~El-Khadra,$^d$ 
E.D.~Freeland,$^{c,d}$ 
E.~G\'amiz,$^{a}$ 
Steven~Gottlieb,$^{f,g}$ 
U.M.~Heller,$^h$ 
J.E.~Hetrick,$^i$
\speaker{Andreas~S.~Kronfeld},$^a$
J.~Laiho,$^j$ 
L.~Levkova,$^e$ 
P.B.~Mackenzie,$^a$ 
M.B.~Oktay,$^e$ 
J.N.~Simone,$^a$ 
R.~Sugar,$^k$ 
D.~Toussaint,$^b$ 
and
R.S.~Van~de~Water$^l$ \\ \\
$^a$Fermi National Accelerator Laboratory,\hspace*{-0.4em}
    \thanks{Operated by Fermi Research Alliance, LLC, under Contract
    No.~DE-AC02-07CH11359 with the United States Department of Energy.}~
 Batavia, IL, USA \\
$^b$Department of Physics, University of Arizona, Tucson, AZ, USA \\
$^c$Department of Physics, Washington University, St.~Louis, MO, USA \\
$^d$Physics Department, University of Illinois, Urbana, IL, USA \\
$^e$Physics Department, University of Utah, Salt Lake City, UT, USA \\
$^f$Department of Physics, Indiana University, Bloomington, IN, USA \\
$^g$National Center for Supercomputing Applications, 
    University of Illinois, Urbana, IL, USA \\
$^h$American Physical Society, One Research Road, Ridge, NY, USA \\
$^i$Physics Department, University of the Pacific, Stockton, CA, USA \\
$^j$SUPA, Department of Physics and Astronomy, 
    University of Glasgow, Glasgow, UK \\
$^k$Department of Physics, University of California, Santa Barbara, CA, USA \\
$^l$Department of Physics, Brookhaven National Laboratory,\hspace*{-0.4em}
    \thanks{Operated by Brookhaven Science Associates, LLC, under 
    Contract No.\ DE-AC02-98CH10886 with the United States Department 
    of Energy.}~
  Upton, NY, USA \\
E-mail: \email{ask@fnal.gov}}
        
\author{Fermilab Lattice and MILC Collaborations}

\abstract{We present an update of our calculation of the form factor 
for $\bar{B}\to D^*l\bar{\nu}$ at zero recoil, with higher statistics 
and finer lattices.
As before, we use the Fermilab action for $b$ and $c$ quarks, the 
asqtad staggered action for light valence quarks, and the MILC 
ensembles for gluons and light quarks (L\"uscher-Weisz married to 2+1 
rooted staggered sea quarks).
In this update, we have reduced the total uncertainty on 
$\mathcal{F}(1)$ from 2.6\% to 1.7\%.

At \textit{Lattice2010} we presented a still-blinded result, but this 
writeup includes the unblinded result from the September 2010 CKM 
workshop.}

\FullConference{The XXVIII International Symposium on 
        Lattice Field Theory, Lattice2010 \\ 
		June 14--19, 2010 \\
		Villasimius, Italy}

\begin{document}

\section{Introduction}

The $Wbc$ vertex is proportional to the coupling $\Vcb$, which is 
an element of the Cabibbo~\cite{Cabibbo:1963yz} Kobayashi-Maskawa~\cite{Kobayashi:1973fv}
(CKM) matrix.
Along with the quark masses, it represents the observable 
part of the quarks' coupling to the Higgs sector and is, thus,
a fundamental part of particle physics.
The CKM matrix has four free parameters, and it is convenient to choose 
one of them to be (essentially) $|\Vcb|$.
Consequently, $|\Vcb|$ appears throughout flavor 
physics~\cite{Antonelli:2009ws}. 

$|\Vcb|$ is determined from semileptonic decays $\bar{B}\to X_cl\bar{\nu}$, 
where $X_c$ denotes a charmed final state.
In \emph{exclusive} decays, $X_c$ is a $D$ or $D^*$ meson, and the 
decay amplitudes can be written 
\begin{eqnarray}
    \langle D(v_D)| \mathcal{V}^\mu|\bar{B}(v_B)\rangle
        &\!=\!&\sqrt{M_B M_{D}}\, \left[
            (v_B + v_D)^\mu h_+(w) + (v_B-v_D)^\mu h_-(w) \right], \\
    \langle D^*(v_D,\alpha)|\mathcal{V}^\mu|\bar{B}(v_B)\rangle 
        &\!=\!&\sqrt{M_BM_{D^*}}\,
            \varepsilon^{\mu\nu}_{\hphantom{\mu\nu}\rho\sigma} 
            \bar{\epsilon}^{(\alpha)}_\nu v^\rho_B v^\sigma_{D^*} h_V(w), \\ 
    \langle D^*(v_D,\alpha)|\mathcal{A}^\mu|\bar{B}(v_B)\rangle 
        &\!=\!&i \sqrt{M_BM_{D^*}}\,\bar{\epsilon}_\nu^{(\alpha)}\!\left\{
            g^{\nu\mu}(1+w)h_{A_1}(w) - 
            v^\nu_B[v^\mu_Bh_{A_2}(w) + v^\mu_{D^*}h_{A_3}(w)]\right\},
        \hspace*{2em}
\end{eqnarray}
where $\epsilon^{(\alpha)}$ is the $D^*$ polarization vector,
$v_B$ and $v_{D^{(*)}}$ denote the mesons' 4-velocities, 
and $w=v_B\cdot v_{D^{(*)}}$ is related to the invariant mass 
of the $l\nu$ pair, $q^2=M_B^2+M_{D^{(*)}}^2-2wM_BM_{D^{(*)}}$.
The form factors $h_\pm$, $h_V$, and $h_{A_i}$ ($i=1,2,3$) enjoy simple
heavy-quark limits and are linear combinations of the form factors
$f_\pm$, $V$, and ${A_i}$ used in other semileptonic decays.

The differential decay distributions are 
\begin{eqnarray}
    \frac{d\Gamma(\bar{B}\to Dl\bar{\nu})}{dw} & = &
        \frac{G^2_F}{48\pi^3} m^3_{D}   (M_B+M_{D})^2  (w^2-1)^{3/2} 
        |V_{cb}|^2 |\mathcal{G}(w)|^2, \\
    \frac{d\Gamma(\bar{B}\to D^*l\bar{\nu})}{dw} & = &
        \frac{G^2_F}{4\pi^3} m^3_{D^*} (M_B-M_{D^*})^2 (w^2-1)^{1/2}
            |V_{cb}|^2 \chi(w)|\mathcal{F}(w)|^2,
\end{eqnarray}
neglecting the charged lepton and neutrino masses.
The physical combinations of form factors are
\begin{eqnarray}
    \mathcal{G}(w) & = & h_+(w) - \frac{M_B-M_D}{M_B+M_D} h_-(w)
        = \frac{2\sqrt{M_BM_D}}{M_B+M_D} f_+(q^2), 
    \label{eq:G(w)} \\ 
    \mathcal{F}(w) & = & h_{A_1}(w)
        \frac{1+w}{2}\sqrt{\frac{H^2_0(w)+H^2_+(w)+H^2_-(w)}{3\chi(w)}} 
        \to h_{A_1}(1) ,
    \label{eq:F(w)} 
\end{eqnarray}
where the zero-recoil ($w\to1$) limit of $\mathcal{F}$ is shown.
The function $\chi(w)$ is chosen so that the square root in 
Eq.~(\ref{eq:F(w)}) collapses to 1 if $h_V=h_{A_3}=h_{A_1}$ and 
$h_{A_2}=0$, as in the heavy-quark limit without 
radiative corrections.
Expressions for $H_\pm(w)$, $H_0(w)$, and $\chi(w)$ can be found in 
Ref.~\cite{Antonelli:2009ws}.

The messy formula for $\mathcal{F}(w)$ indicates the advantage of the 
zero-recoil limit for 
$\bar{B}\to D^*l\bar\nu$: one must compute only $h_{A_1}(1)$, not four 
functions.
In addition, the heavy-quark flavor symmetry is larger when 
$v_{D^*}=v_B$, and Luke's theorem applies.
For determining $|\Vcb|$, the key aspect of Luke's theorem is that it 
helps control systematic errors.
In particular, in lattice gauge theories that respect heavy-quark 
symmetry, one can compute $h_{A_1}(1)$ with heavy-quark discretization 
errors that are formally $\bar{\Lambda}/m_Q$ times smaller than those 
of $h_{A_1}(w)$, $w\neq1$, or those of $\mathcal{G}(w)$ even at $w=1$.


Here we focus on $\bar{B}\to D^*l\bar{\nu}$ at zero recoil, describing 
our calculations of $\Fone=h_{A_1}(1)$.
Starting in 2001, experimental determinations of $|\Vcb|$ used a quenched 
calculation~\cite{Hashimoto:2001nb} 
\begin{equation}
    \Fone = 0.913^{+0.024}_{-0.017}\pm 0.016^{+0.003}_{-0.014}
        {}^{+0.000}_{-0.016}{}^{+0.006}_{-0.014},
    \label{eq:quenched}
\end{equation}
where the errors stem, respectively, from statistics, matching lattice 
gauge theory to QCD, lattice-spacing dependence, chiral extrapolation, 
and the quenched approximation.
\pagebreak
A notable feature of Eq.~(\ref{eq:quenched}) is that an estimate of the 
error associated with quenching has been made.
Nevertheless, it is necessary to incorporate the light- and 
strange-quark sea.
The first calculation with 2+1 flavors of sea quarks 
obtained~\cite{Bernard:2008dn}
\begin{equation}
    \Fone = 0.921\pm 0.013\pm 0.008\pm 0.008\pm 0.014\pm 0.003\pm 
        0.006\pm 0.004,
    \label{eq:2+1}
\end{equation}
where, now, the errors stem from statistics, the $g_{D^*D\pi}$ 
coupling, chiral extrapolation, discretization errors, matching, 
and two tuning errors.
(The catch-phrases for the errors do not have exactly the same meaning 
in Refs.~\cite{Hashimoto:2001nb,Bernard:2008dn}; for example, the 
$g_{D^*D\pi}$ error in Eq.~(\ref{eq:quenched}) is incorporated into the 
chiral-extrapolation error.)
This paper presents an update of the 2+1-flavor calculation,
with mostly the same ingredients, but with higher statistics and 
without the second of the tuning errors.

The new data set is shown in Table~\ref{tab:ens}, based as before on 
the MILC ensembles~\cite{Bazavov:2009bb} 
with the L\"uscher-Weisz gauge action~\cite{Luscher:1984xn}, 
with the $g^2N_c$~\cite{Luscher:1985zq} but not $g^2N_f$ 
corrections~\cite{Hao:2007iz}, 
and the asqtad-improved~\cite{Lepage:1998vj} rooted staggered 
determinant for the sea quarks.
\begin{table}[t]
\caption[tab:ens]{Parameters of the MILC ensembles used for
    heavy-quark physics. 
    Here $C$ denotes the number of configurations in each ensemble;
    $(m'_l,m'_s)$ the asqtad sea-quark masses;
    $m_q$ the asqtad valence masses;
    $\kappa$~and \cSW\ the hopping parameter and clover coupling of the 
    heavy quark.
    Standard nicknames for the lattice spacings are noted 
    ($a\approx0.045$~fm is ``ultrafine'').
    Data are being generated on all ensembles for all $m_q$ inside the 
    $\{\cdots\}$, but the present analysis uses at most two, namely
    $m_q=m'_l$ and $m_q=0.4m'_s$.}
\label{tab:ens}
\begin{tabular}{cr@{$^3\times$}lrr@{,}lr@{, }lccl}
    \hline\hline
    $a$ (fm) & \multicolumn{2}{c}{Lattice} & $C$\hspace*{0.5em} & $(am'_l$&$am'_s)$ & 
    \multicolumn{2}{c}{$m_q$} & \kbot & \kch & ~~\cSW \\
    \hline 
    $\approx0.15$ & 16&48 & 596 & (0.0290&0.0484) & \{0.0484&0.0453, \\
    {\small medium} & 16&48 & 640 & (0.0194&0.0484) &   0.0421&0.0290, \\
    {\small coarse} & 16&48 & 631 & (0.0097&0.0484) & 0.0194&0.0097, &
        0.0781 & 0.1218 & 1.570 \\
            & 20&48 & 603 & (0.0048&0.0484) & 0.0068&0.0048\} \\
    \hline 
    $\approx0.12$ & 20&64 & 2052 & (0.02&0.05) & \{0.05&0.03, & 
        0.0918 & 0.1259 & 1.525 \\
    {\small coarse}        & 20&64 & 2259 & (0.01&0.05) & 0.0415&0.0349, & 
        0.0901 & 0.1254 & 1.531 \\
            & 20&64 & 2110 & (0.007&0.05) & 0.02&0.01, & 
        0.0901 & 0.1254 & 1.530 \\
            & 24&64 & 2099 & (0.005&0.05) & 0.007&0.005\} & 
        0.0901 & 0.1254 & 1.530 \\
    \hline 
    $\approx0.09$ & 28&96 & 1996 & (0.0124&0.031) & \{0.031&0.0261, & 
        0.0982 & 0.1277 & 1.473 \\
    {\small fine}          & 28&96 & 1946 & (0.0062&0.031) & \multicolumn{2}{c}{0.0124,} & 
        0.0979 & 0.1276 & 1.476 \\
            & 32&96 & 983 & \hspace*{-6pt}(0.00465&0.031) & 0.0093&0.0062, & 
        0.0977 & 0.1275 & 1.476 \\
            & 40&96 & 1015 & (0.0031&0.031) & 0.0047&0.0031\} & 
        0.0976 & 0.1275 & 1.478 \\
    \hline 
    $\approx0.06$ & 48&144 & 668 & (0.0072&0.018) & \{0.0188&0.0160, & 
        0.1052 & 0.1296 & 1.4276\hspace*{-3pt} \\
    {\small superfine} & 48&144 & 668 & (0.0036&0.018) & \multicolumn{2}{c}{0.0072,} & 
        0.1052 & 0.1296 & 1.4287\hspace*{-3pt} \\
        & 56&144 & 800 & (0.0025&0.018) & 0.0054&0.0036, \\
      & 64&144 & 826 & (0.0018&0.018) & 0.0025&0.0018\} \\
    \hline
    $\approx0.045$ & 64&192 & 860 & (0.0028&0.014) & 
        \multicolumn{3}{l}{\{0.014, 0.0056, 0.0028\}} & & \\
    \hline\hline
\end{tabular}
\end{table}
For the valence quarks, we use the asqtad action for the light quark 
and the Fermilab interpretation~\cite{ElKhadra:1996mp} 
of the clover action~\cite{Sheikholeslami:1985ij} for the heavy quark.
In this report, we use all ensembles in Table~\ref{tab:ens} with 
entries for the heavy-quark couplings (\kbot, \kch, and \cSW), 
\emph{except} the fine $32^3\times96$ lattice. 
These data are being generated as part of a broad program of 
heavy-quark physics, including other semileptonic 
decays~\cite{ElviraJon} and neutral-meson mixing and decay 
constants~\cite{ChrisB}.

Improvements to \Fone\ are timely~\cite{Antonelli:2009ws}, because the 
values of $|\Vcb|$ that follow from inclusive decays are in a 
$2.2\sigma$ tension with those that follow from Eq.~(\ref{eq:2+1}) and also 
from $\bar{B}\to Dl\bar{\nu}$ and $\mathcal{G}(1)$ \cite{Okamoto:2004xg}.
The result described below is but one aspect of a resolution of the 
discrepancy.
Others include a re-examination of the extrapolation to zero recoil,
unquenched lattice-QCD calculations at $w\neq1$, 
lattice-QCD calculations by other groups~\cite{de Divitiis:2007ui}, 
and the incorporation of higher-order corrections to the 
inclusive decay expressions.

In Sec.~\ref{sec:data}, we discuss details of the data and of the data 
analysis.
Because the value of $\Fone$ has been studied so much in the past, any 
new analysis could be influenced in subtle human ways.
To circumvent any such bias, we hide the numerical value of $\Fone$ via 
an offset in the matching factor $\rho_{A^{\mathrm{cb}}}$, 
explained in Sec.~\ref{sec:blind}.
We present our preliminary results, with all sources of uncertainty 
estimated, in Sec.~\ref{sec:result}.
We include the unblinded value here, which was revealed after 
\emph{Lattice 2010} but before these proceedings.

\section{Data analysis}
\label{sec:data}

As in Ref.~\cite{Bernard:2008dn}, we aim for the direct double-ratio 
\begin{equation}
    \mathcal{R}_{A_1} = \frac{
        \langle    D^*|\bar{c}\gamma_j\gamma_5b|\bar{B}\rangle 
        \langle\bar{B}|\bar{b}\gamma_j\gamma_5c|D^*\rangle}{%
        \langle    D^*|\bar{c}\gamma_4 c|D^*\rangle
        \langle\bar{B}|\bar{b}\gamma_4 b|\bar{B}\rangle} =
        \left|h_{A_1}(1)\right|^2,
    \label{eq:RA1}
\end{equation}
where the expressions here are all in (continuum) QCD.
To this end, we use lattice gauge theory to compute the three-point 
correlation functions 
\begin{eqnarray}
    C^{B\to D^*}(t_i,t_s,t_f) & = & \sum_{\bm{x},\bm{y}}
        \langle 0|{\cal O}_{D^*}(\bm{x},t_f)
        \overline{\Psi}_{\mathrm{c}}\gamma_j\gamma_5\Psi_{\mathrm{b}}(\bm{y},t_s)
        {\cal O}^\dag_B(\bm{0},t_i)| 0\rangle, \\ 
    C^{B\to B}(t_i,t_s,t_f) & = & \sum_{\bm{x},\bm{y}}
        \langle 0|{\cal O}_{B}(\bm{x},t_f)
        \overline{\Psi}_{\mathrm{b}}\gamma_4\Psi_{\mathrm{b}}(\bm{y},t_s)
        {\cal O}^\dag_B(\bm{0},t_i)| 0\rangle, \\
    C^{D^*\to D^*}(t_i,t_s,t_f) & = & \sum_{\bm{x},\bm{y}}
        \langle 0|{\cal O}_{D^*}(\bm{x},t_f)
        \overline{\Psi}_{\mathrm{c}}\gamma_4\Psi_{\mathrm{c}}(\bm{y},t_s)
        {\cal O}^\dag_{D^*}(\bm{0},t_i)|0\rangle,
\end{eqnarray}
where $\mathcal{O}_B$ and $\mathcal{O}_{D^*}$ are interpolating 
operators coupling to the $B$ and $D^*$ mesons,
and $\overline{\Psi}_{\mathrm{c}}\gamma_\mu\Psi_{\mathrm{b}}$ and
$\overline{\Psi}_{\mathrm{c}}\gamma_\mu\gamma_5\Psi_{\mathrm{b}}$ are
improved currents~%
\cite{ElKhadra:1996mp,Hashimoto:2001nb,Harada:2001fj,Bernard:2008dn}.
Then the lattice ratio
\begin{equation}
    R_{A_1}(t) = \frac{C^{B\to D^*}(0,t,T)
        C^{D^*\to B}(0,t,T)}{C^{D^*\to D^*}(0,t,T)C^{B\to B}(0,t,T)} 
\end{equation}
should reach a plateau for a range of $t$, $T\gg t\gg 1$.
The relationship between the plateau value of $R_{A_1}^{1/2}$ and 
$h_{A_1}(1)$ is discussed in Sec.~\ref{sec:blind}. 

With staggered fermions, $\mathcal{O}_B$ and $\mathcal{O}_{D^*}$ 
couple to both parities, and three-point correlation functions have 
four distinct contributions:
\begin{eqnarray}
    C^{X\to Y}(0,t,T) & = & \sum_{k=0}\sum_{\ell=0}
        (-1)^{kt}(-1)^{\ell(T-t)}
        A_{\ell k}e^{-M^{(k)}_{X}t}e^{-M^{(\ell)}_{Y}(T-t)} \\
    & = & A^{X\to Y}_{00}e^{-M_X t - M_Y(T-t)} + 
        (-1)^{T-t} A^{X\to Y}_{01}e^{-M_X t-M_Y'(T-t)} + \nonumber \\
    &  & (-1)^t A^{X\to Y}_{10} e^{-M_X' t -M_Y(T-t)} +
        (-1)^T A^{X\to Y}_{11} e^{-M_X't-M_Y'(T-t)} + \cdots
\end{eqnarray}
with time-dependent factors of $-1$ associated with the states of 
undesired parity.
To reduce the magnitude of the oscillating components, we form the 
combination~\cite{Bernard:2008dn}
\pagebreak
\begin{equation}
    \bar{R}_{A_1}(0,t,T) = \case{1}{2} R_{A_1}(0,t,T) + 
        \case{1}{4} R_{A_1}(0,t,T+1)  + \case{1}{4}R_{A_1}(0,t+1,T+1),
\end{equation}
which should tend more quickly to a plateau.
The key here is to have $t_f=T$ and $T+1$.

\begin{figure}[b]
    \includegraphics[width=0.3\textwidth]{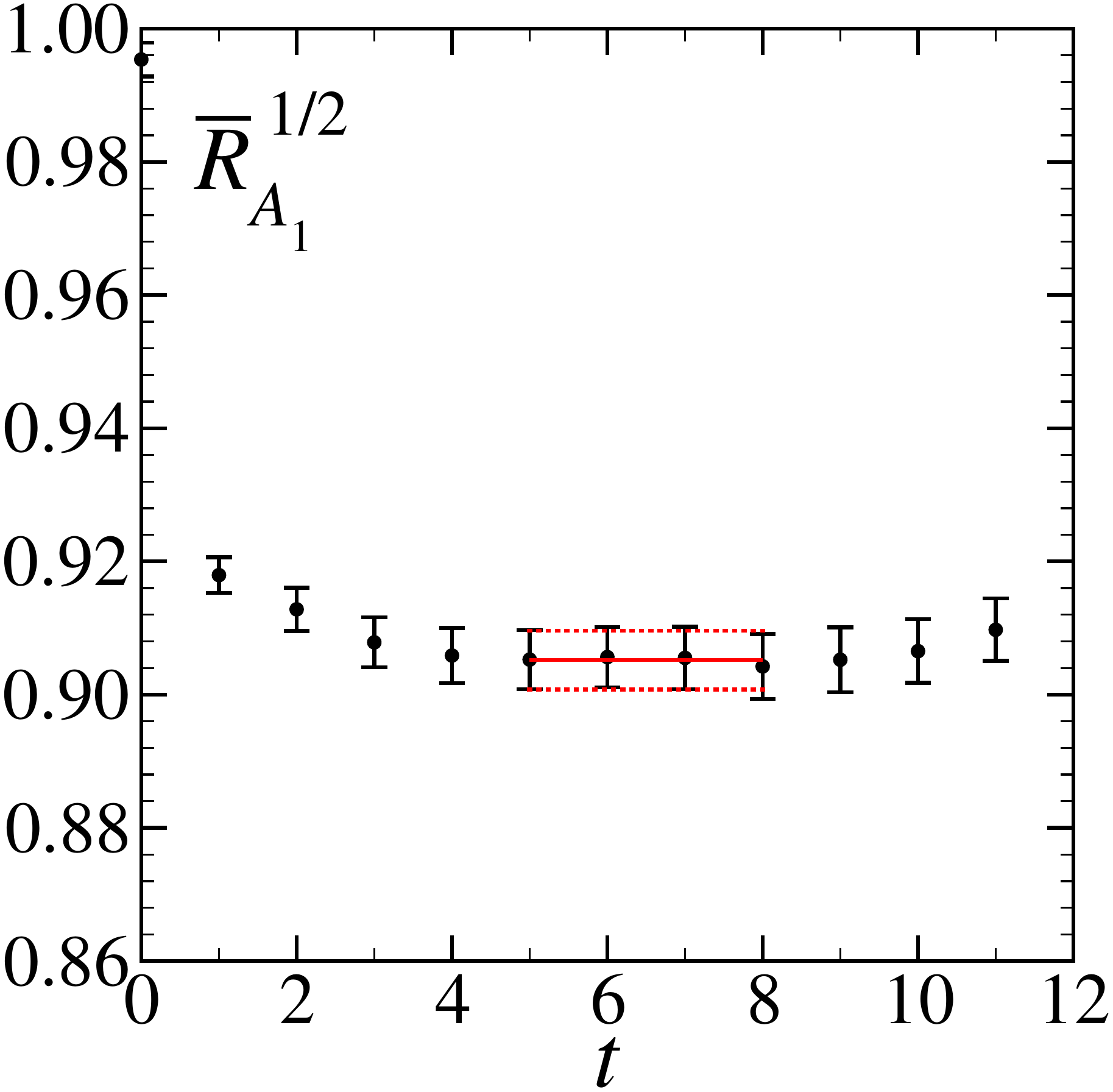}\hfill
    \includegraphics[width=0.3\textwidth]{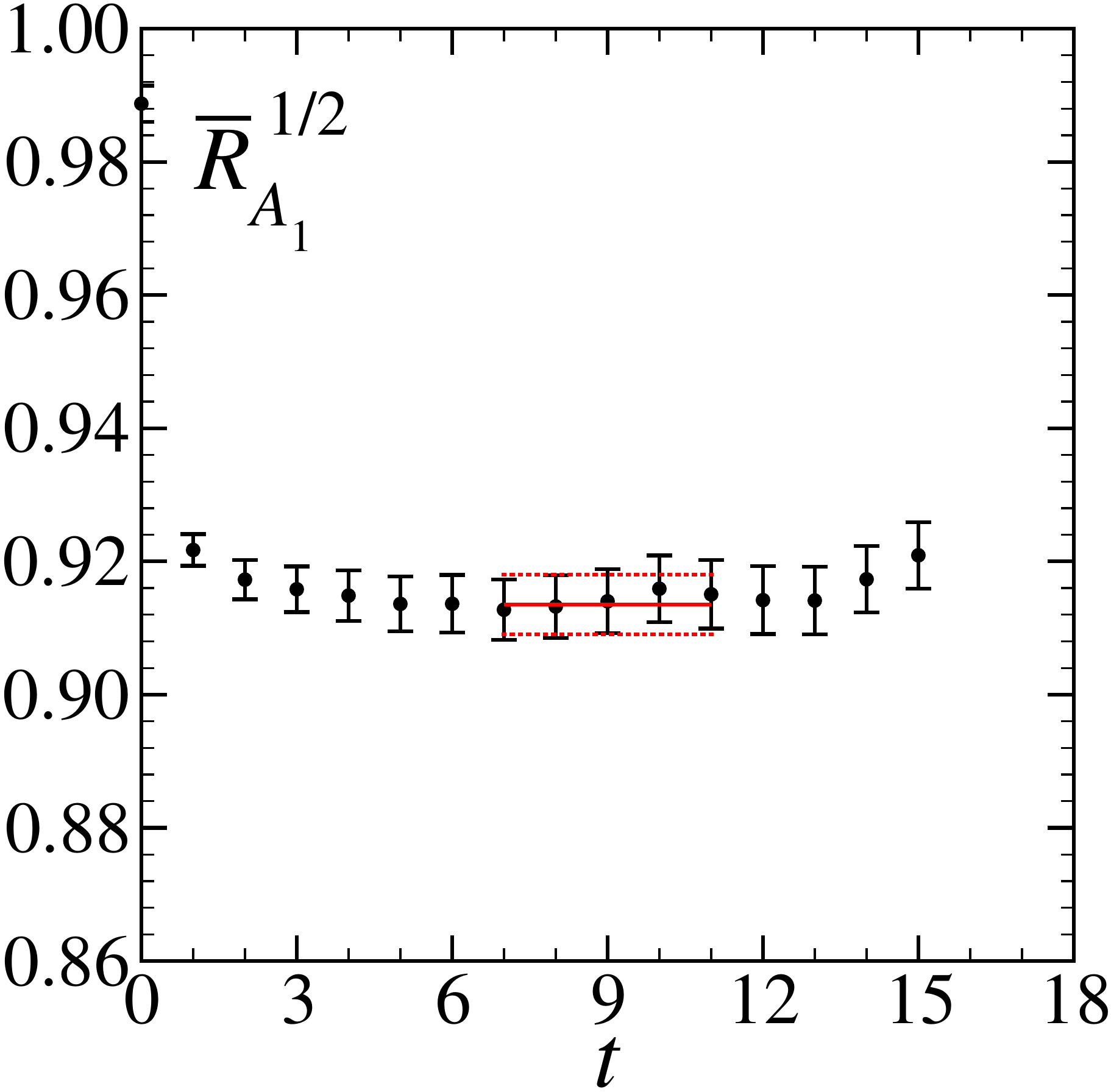}\hfill
    \includegraphics[width=0.3\textwidth]{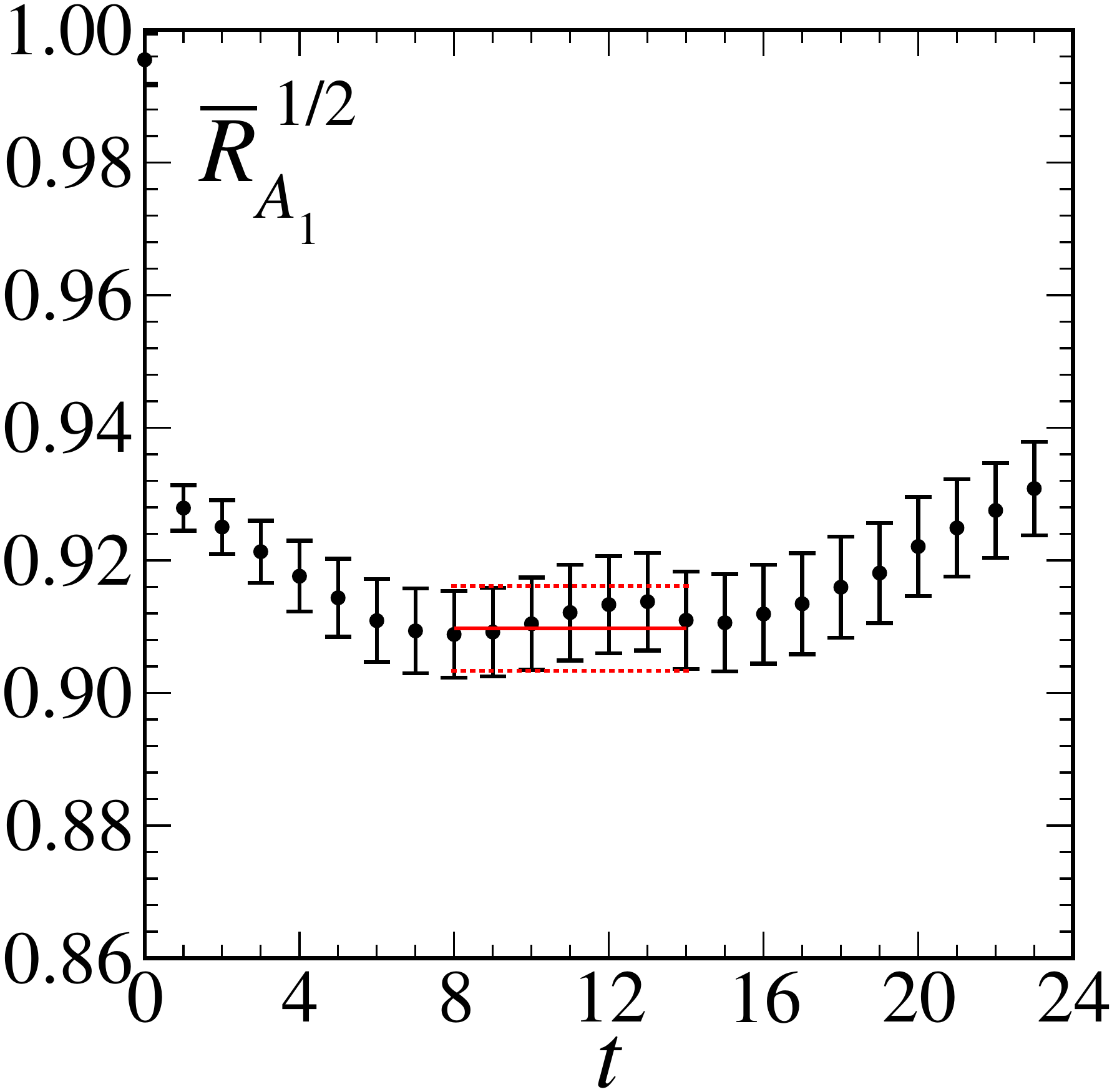}
    \caption[fig:RA1]{Ratio combination $\bar{R}_{A_1}^{1/2}(0,t,T)$ \emph{vs.}\ $t$
        with $m_q=m'_l=0.2m'_s$.
        From left to right: 
        the coarse ensemble with $T=12$ and $(am'_l,am'_s)=(0.01,0.05)$;
        the fine ensemble with $T=17$ and $(am'_l,am'_s)=(0.0062,0.031)$;
        the superfine ensemble with $T=24$ and $(am'_l,am'_s)=(0.0036,0.018)$.}
    \label{fig:RA1}
\end{figure}

The correlation functions and their ratios are analyzed 
for two light valence quark masses per ensemble, namely, 
$m_q=m'_l$ and $m_q=0.4m'_s$ (or the single $m_q$ when $m'_l=0.4m'_s$).
The choice of a fixed $m_q$, here $0.4m'_s$, for all $m'_l$ matches, by 
design, our plans for the ultrafine lattice ($a\approx0.045$~fm), to 
anchor future analyses even closer to the continuum limit.
Typical plateaus are shown in Fig.~\ref{fig:RA1}
for a coarse, a fine, and a superfine ensemble.
As one can see, the plateau in $\bar{R}_{A_1}$ emerges readily, and the 
statistical errors are 1\% or smaller.

\section{Matching, blinding, and discretization effects}
\label{sec:blind}

The ratio combination $\bar{R}_{A_1}$ tends to a ratio of matrix 
elements like $\mathcal{R}_{A_1}$ in Eq.~(\ref{eq:RA1}) but with 
lattice currents. 
Each current must be multipled by a matching factor $Z_A$ or $Z_V$, 
defined nonperturbatively in Ref.~\cite{Harada:2001fj}.
The lattice ratio $R_{A_1}$ must, therefore, be multiplied by a matching ratio
\begin{equation}
    \rho_{A^{\mathrm{cb}}}^2 = 
        Z_{A^{\mathrm{cb}}}^2/Z_{V^{\mathrm{cc}}}Z_{V^{\mathrm{bb}}}.
\end{equation}
A subset of the collaboration has computed $\rho_{A^{\mathrm{cb}}}$ in 
the one-loop approximation.
The result is very close to unity, but the deviation is, or could be, 
comparable to $h_{A_1}(1)-1$.
Our numerical analysis replaces $\rho_{A^{\mathrm{cb}}}$ with
$F_{\mathrm{blind}}\rho_{A^{\mathrm{cb}}}$, where the 
\emph{blinding factor} $F_{\mathrm{blind}}$ is again close to unity, 
but known only to those engaged in the one-loop calculation.
In this way, choices of fitting ranges, etc., cannot be influenced by a 
human desire to (dis)agree with results for \Fone\ already in the 
literature. 

The HQET-Symanzik formalism used to define the $Z_J$ can also be used 
to control and suppress cutoff dependence~\cite{Kronfeld:2000ck,Harada:2001fj}.
In the general case, several operators---both corrections to the 
current and insertions of the effective Lagrangian---generate cutoff 
effects.
For details, see, \emph{e.g.}, the discussion of Eq.~(2.40) in 
Ref.~\cite{Harada:2001fj}.
For zero recoil, $v_{D^*}=v_B$, and the heavy-quark flavor 
symmetry enlarges from $\mathrm{U}(1)\times\mathrm{U}(1)$ to 
$\mathrm{SU}(2)$.
The leading discretization errors drop out, and the remainder can be 
found by applying the formulas of Ref.~\cite{Kronfeld:2000ck} to 
$\bar{R}_{A_1}^{1/2}$ and~$h_{A_1}(1)$. 
One finds
\begin{equation}
    \rho_{A^{\mathrm{cb}}} \bar{R}_{A_1}^{1/2} = h_{A_1}(1) + 
        \mathrm{O}(\alpha_s a\bar{\Lambda}^2/m_{\mathrm{c}}) +
        \mathrm{O}(\alpha_s a^2 \bar{\Lambda}^2) +
        \mathrm{O}(\alpha_s^2),
    \label{eq:discret}
\end{equation}
where the last error acknowledges the one-loop calculation of
$\rho_{A^{\mathrm{cb}}}$.
\pagebreak
A study of the asymptotic behavior of Fermilab actions provides 
a reasonable guide to the dependence on $m_Qa$ of the corrections.
We see in our data little dependence on the lattice spacing, in accord 
with Eq.~(\ref{eq:discret}).

\section{Preliminary result}
\label{sec:result}

\begin{figure}[b]
    \includegraphics[width=0.45\textwidth]{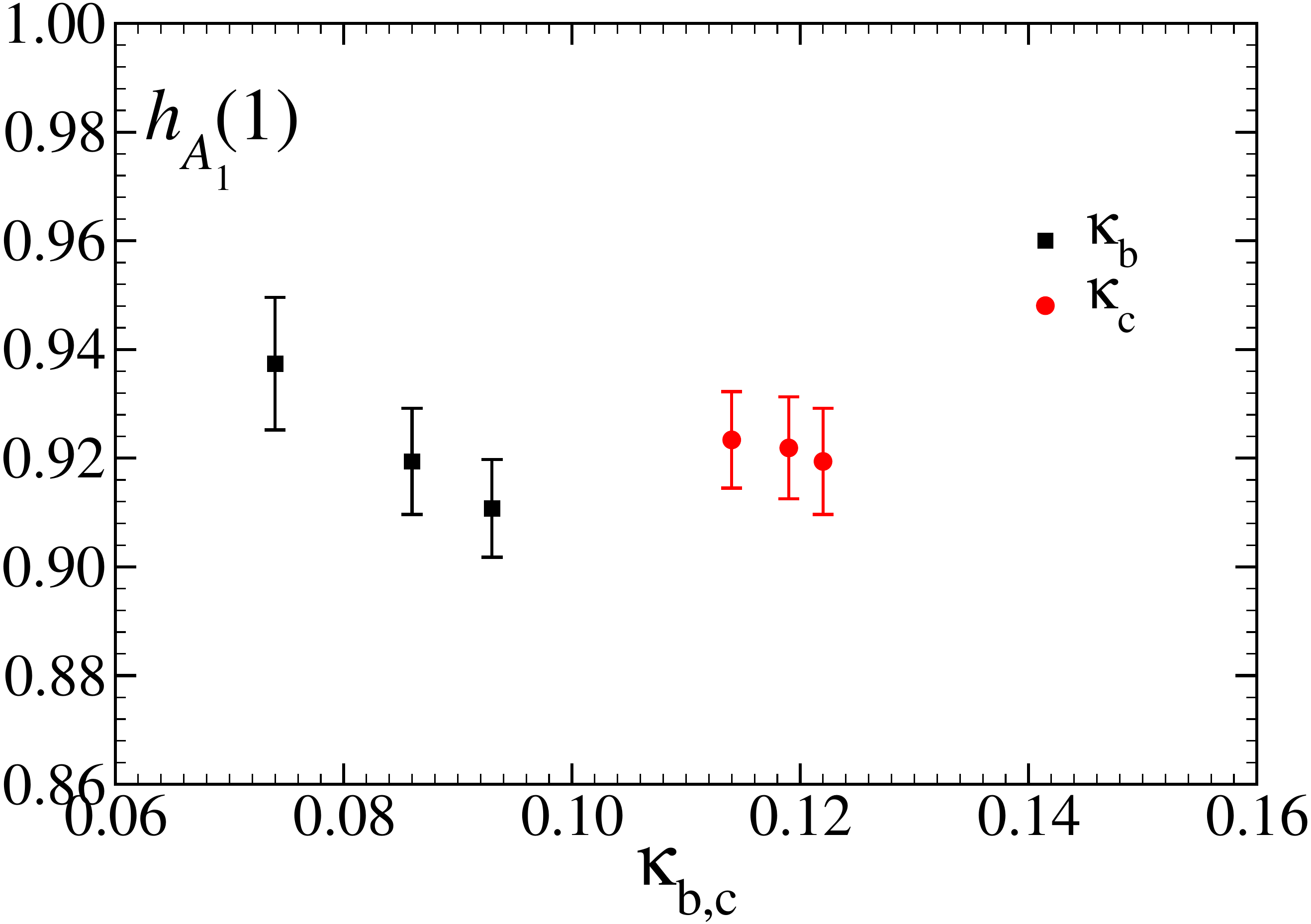}\hfill
    \includegraphics[width=0.45\textwidth]{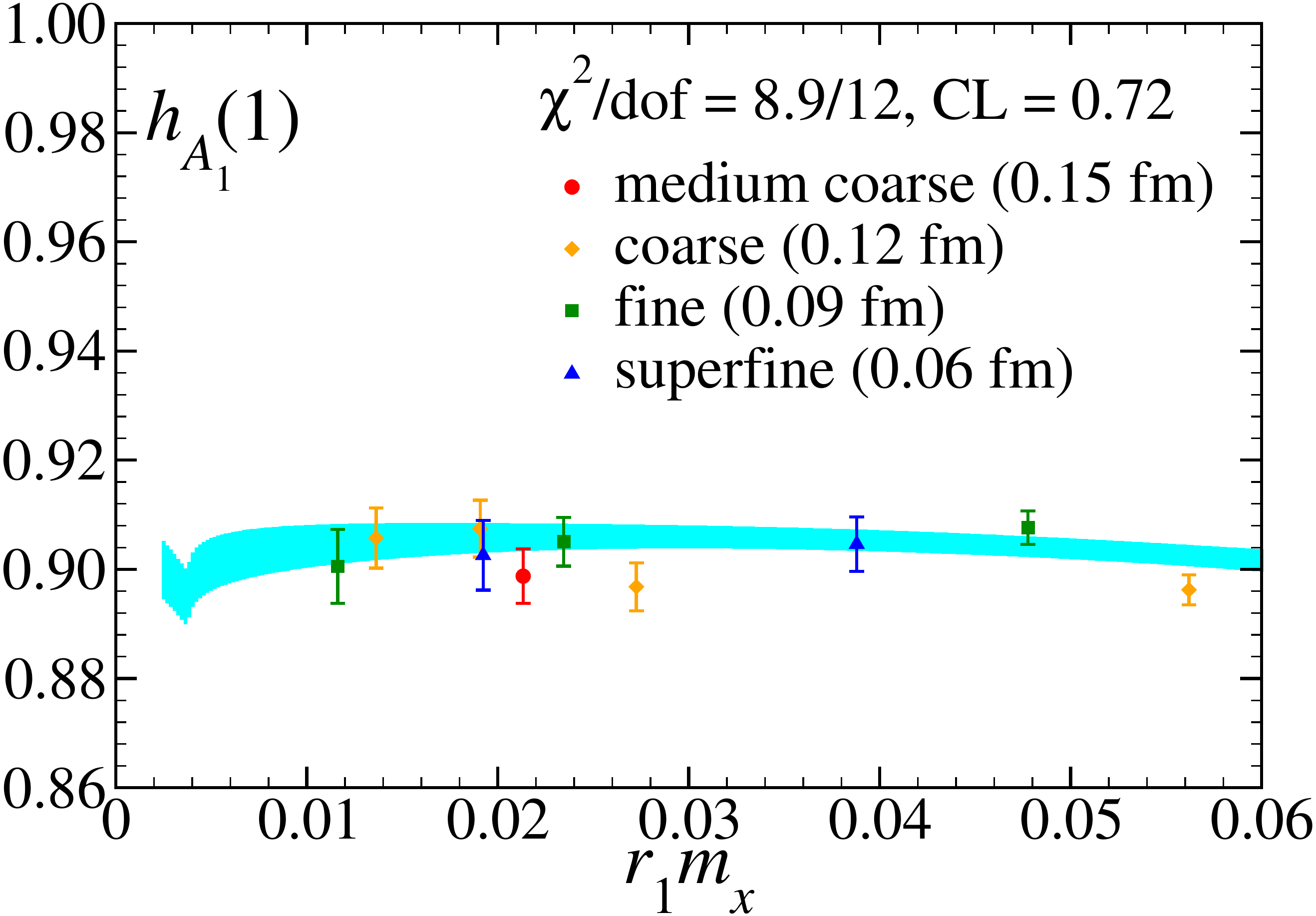}
    \caption[fig:other]{Left: dependence of $h_{A_1}(1)$ on the 
    heavy-quark hopping parameters (with data of Ref.~\cite{Bernard:2008dn}).
    Right: chiral extrapolation showing only points with $m_q=m_l$ and 
    a fit to all data.}
    \label{fig:systematics}
\end{figure}

Figure~\ref{fig:systematics} provides a glimpse into our systematic 
error analysis, which closely follows Ref.~\cite{Bernard:2008dn}.
We use our previous study of heavy-quark-mass dependence to fine-tune 
\emph{a posteriori} the hopping parameters and to assess the tuning 
errors.
We fit the light-quark mass dependence to one-loop chiral perturbation 
theory, suitably modified for staggered quarks~\cite{Laiho:2005ue}.
The cusp is a necessary, physical effect that appears because the
$D\pi$ threshold sinks below the $D^*$ mass.

With the blinding factor in place, we find
\begin{equation}
    F_{\mathrm{blind}}\mathcal{F}(1) = 0.8949 \pm 0.0051 \pm 0.0088
        \pm 0.0072 \pm 0.0093 \pm 0.0030 \pm 0.0050,
    \label{eq:blinded}
\end{equation}
where the errors again stem from statistics, the $g_{D^*D\pi}$ 
coupling, chiral extrapolation, discretization errors, matching, 
and tuning \kch\ and \kbot.
To show how the errors have been reduced, it helps to scale this result 
to the old central value ($F_F$ is the needed \emph{ad hoc} factor):
\begin{eqnarray}
    \mathcal{F}(1)    & = & 
        0.921(13)(8)(8)(14)(3)(6)(4)~\textrm{\cite{Bernard:2008dn}}, \\
    F_F\mathcal{F}(1) & = & 
        0.921(05)(9)(7)(10)(3)(5)~~~~~~\,\textrm{[this work]}.
\end{eqnarray}
The higher statistics and wider scope of this dataset has reduced the 
statistical error with $C^{-1/2}$.
The quoted heavy-quark discretization error is smaller, because with 
the superfine data we can move beyond pure power counting and combine 
the (lack of) trend in the data with the detailed theory of cutoff 
effects~\cite{Kronfeld:2000ck}.
After Lattice 2010, we continued to examine the heavy-quark 
discretization and $\kappa$-tunings errors, reducing them somewhat, and 
the chiral-extrapolation error, increasing it somewhat.
For the 2010 \emph{Workshop on the CKM Unitarity Triangle}, 
we removed the blinding factor, finding~\cite{Mackenzie:2010CKM}:
\begin{equation}
    \mathcal{F}(1) = 0.9077(51)(88)(84)(90)(30)(33).
    \label{eq:CKM2010}
\end{equation}
This result reduces the tension with $|\Vcb|$ from inclusive decays 
to~$1.6\sigma$.

Computations for this work were carried out in part on facilities of
the USQCD Collaboration, which are funded by the Office of Science of
the U.S. Department of Energy.
This work was supported in part by the U.S. Department of Energy under 
Grants No.~DE-FC02-06ER41446 (C.D., L.L., M.B.O),
No.~DE-FG02-91ER40661 (S.G.), No.~DE-FG02-91ER40677 (C.M.B., A.X.K., E.D.F.),
No.~DE-FG02-91ER40628 (C.B, E.D.F.), No.~DE-FG02-04ER-41298 (D.T.); 
the National Science Foundation under Grants No.~PHY-0555243, No.~PHY-0757333,
No.~PHY-0703296 (C.D., L.L., M.B.O), No.~PHY-0757035 (R.S.), 
No.~PHY-0704171 (J.E.H.) and No.~PHY-0555235 (E.D.F.).
C.M.B. was supported in part by a Fermilab Fellowship in Theoretical 
Physics
and by the Visiting Scholars Program of Universities Research Association, Inc.
R.S.V. acknowledges support from BNL via the Goldhaber Distinguished Fellowship.

\end{document}